\documentclass{aa}  

\usepackage{graphicx}
\usepackage{txfonts}
\usepackage{lipsum}
\usepackage{subcaption}         
\usepackage{lscape}             
\usepackage{placeins}           
                                
\begin{document}

   \title{Reinterpreting the sunward electron deficit: Implications for solar wind acceleration and core population formation}


%


   \author{Z. Nemeth\thanks{Corresponding author: nemeth.zoltan@wigner.hun-ren.hu}}

   \institute{HUN-REN Wigner Research Centre for Physics, Konkoly-Thege Miklós út 29-33., Budapest H-1121, Hungary}

   \date{}

 
  \abstract
   {}
   {The paper re-evaluates the relationship between the observed sunward electron cutoff energy and the depth of the Sun’s global electrostatic potential. It investigates whether considering the effects of moving particle traps formed by magnetic fluctuations could provide an alternative explanation for the observed electron deficit.} 
   {The study assesses the effects of the traps on the electron velocity distribution and derives mathematical expressions for key parameters. It also estimates the parameters using observations of the Parker Solar Probe. It investigates how the nature of the trap-forming magnetic disturbances influences electron behavior, and compares the trapping efficiency of magnetic field intensity variations with that of magnetic switchbacks.}
   {The effective potential governing electron motion exhibits fundamentally different behaviors depending on the distance from the Sun. Within a critical distance, fluctuations cannot yet establish stable traps. Beyond the critical distance, electrons are captured and carried away by moving magnetic traps. Consequently, the observed electron cutoff energy characterizes only the potential drop within the critical distance. The true potential can be significantly deeper, scaled by a factor related to the ratio of the measurement site's distance to the critical distance. The critical distance is approximately 30–50 solar radii. The influence of the traps leads to an upward adjustment of the best fit for the power-law exponent of the radial variation of the potential. Mirror traps would create a velocity-dependent cutoff, whereas switchbacks yield a universal cutoff and generally good agreement with observations.}
   {The Sun's electrostatic potential is a more significant factor in solar wind acceleration than previously interpreted from cutoff data; even the acceleration of the fast solar wind can be accounted for by the potential. Electrons captured by these moving traps contribute to the formation of the solar wind core population.}


   \maketitle
   \nolinenumbers

\section{Introduction}

Recent space missions, especially the unprecedented observations of the
Parker Solar Probe \citep{fox2016}, have enabled us to gain previously
unattainable knowledge about the nascent solar wind. Based on
theoretical considerations, it has long been assumed that the Sun has an
electrostatic field (formed because the Sun's
gravitational field alone cannot retain the more mobile electrons),
which plays a significant role in accelerating solar wind ions \citep{lemaire1971}. Previously, it was not possible to experimentally
investigate this field, but recently, some clues pointing to its
existence have been discovered in the anisotropies of particle
distributions observed near the Sun. Several articles \citep{bercic2021,halekas2020,halekas2021,halekas2022,raouafi2023} report that
some of the sunward moving electrons are missing above a certain energy
level. Since the simplest explanation for the existence of a population
moving toward the Sun is that an attractive potential turned back the
out-moving particles, the observed energy-dependent cutoff may indicate
that above this energy, electrons are able to escape the attractive
potential, i.e., the cutoff energy may characterize the depth of the
potential relative to the zero potential defined in spatial infinity.
Based on the above assumptions, calculating the depth of the potential
from the cutoff energy leads to the conclusion that, except for the slowest solar wind, the electric
potential of the Sun is significantly lower than what would be necessary
to accelerate the solar wind ions to the observed velocities \citep{halekas2022}. This
conclusion (if true) relegates the electric potential to the role of a
minor factor in solar wind acceleration and reopens the burning question
of what process might dominate ion acceleration in the collisionless
plasma of the solar wind.

It is tempting to suggest that we should revert to the fluid model of
solar wind acceleration pioneered by Eugene Parker \citep{parker1958},
especially since \cite{nemeth2025} demonstrated that, under certain
conditions, magnetized plasmas can behave as fluids even in the absence
of particle-particle collisions. However, simple energy considerations
indicate that fluid effects alone cannot fully account for the
acceleration of the solar wind. The pressure gradient can only serve to
concentrate all the energy distributed among the different degrees of
freedom of the particles into a single translational degree of freedom.
(This is the same principle, which also underlies the theoretical limit
of outflow velocity that can be achieved using a de Laval nozzle.)
Simply put, if all particles entering the acceleration region exit on
the other side while no additional energy is introduced above the
thermal energy already present, then acceleration only means the
redistribution of energy between degrees of freedom -- hence the upper
bound described above. The introduction of an electric potential,
particularly one that shifts from attractive to repulsive as an ion
moves away from the Sun, addresses this issue \citep{lemaire2010,maksimovic1997,zouganelis2004}. Not all particles that enter the
process, and contribute energy, will escape the potential well of the
Sun. The portion that escapes takes energy away from the remaining
plasma, which consequently cools down. This aspect of the acceleration
is necessary to allow the observed high speed of the solar wind flow,
and this is what is called into question by the new observations.

To resolve this problem, we can proceed in one of two ways: either we
try to find a new solar wind acceleration process with appropriate
characteristics to replace the electric potential, or we can re-examine
the logic that relates the observed cutoff energy of sunward-moving
electrons to the depth of the potential well. In this paper, we pursue
the second option.

\section{Methods and Results}
What we can say with certainty based on the observations is that, above
a certain energy level, something prevents the outward-moving electrons
from returning to the measurement location. As we saw earlier, one
possibility is that these electrons completely leave the potential well
and disappear into infinity. The question is whether there is another
possibility.

In magnetized plasma, scattering on magnetic fluctuations plays a
decisive role in shaping the motion of particles along field lines.
These interactions determine not only the parallel pressure \citep{chew1956}, but also the fundamental nature of plasma behavior \citep{nemeth2025}.
We can thus envision the entire field line system as a complex network
of shallow magnetic traps, which are bounded by magnetic fluctuations. If the
energy of a particle is sufficiently small, it can be reflected by a
strong magnetic fluctuation or can even bounce back and forth in a
magnetic bottle formed by two significant field variations.
Add to this the effect of electrons continuously losing energy as they
move outward in an attractive potential together with the fact that the
magnetic traps also move away from the Sun on average as they travel
with the solar wind, and another possibility for explaining the observed
cutoff energy begins to emerge.

Solar wind electrons move under the influence of an effective potential $V_{eff}$, which has a time independent main part slowly changing with distance along the field line; and a dynamic part corresponding to moving magnetic field disturbances. The main part comprises the electrostatic potential mentioned earlier, the gravitational potential (insignificant for the electrons), and a contribution from the average (non-fluctuating) part of the magnetic field due to adiabatic cooling.
From the constancy of the first adiabatic invariant and the invariance of the total energy, we have
\begin{equation}
    \frac{1}{2}mv_{\parallel}^2 + e\Phi + \mu B = const
    \label{eq1}
\end{equation}
and thus $\mu B$ contributes to the effective potential, which governs the parallel motion of the electrons, $V_{eff} = e\Phi + \mu B$ here. This equation should be treated with caution, because, although $e$ is a universal constant, $\mu$ is merely a constant of motion (within the adiabatic approximation) and can change from particle to particle. Since $B$ is highly variable, it is easy to see from Eq. \ref{eq1} how fluctuations of the magnetic field magnitude can directly contribute to the potential governing the parallel motion of the electrons. Later, we will see that other types of magnetic disturbances, especially switchbacks \citep{bale2019, kasper2019}, can also produce similar contributions to the effective potential. 

Figure 1 illustrates the general behavior of the effective potential in the presence of a fluctuating magnetic field, where the fluctuations are represented by a simple sinusoidal function. We can identify two radically different regions: closer to the Sun, the effective potential increases monotonically, whereas in the more distant regions, periodic local minima form particle traps. Since the magnetic fluctuations are moving away from the Sun, the local minima are also shifting, and the electrons trapped inside the minima are forced to move together with the traps. The two insets in the figure highlight magnified sections of the effective potential to emphasize the differing behaviors; light blue shading inside the minima in the right-hand inset represents the trapped electrons. From now on, we denote the critical distance separating the two regions with $r_c$.
   \begin{figure}[ht!]
        \centering
        \includegraphics[width=\linewidth]{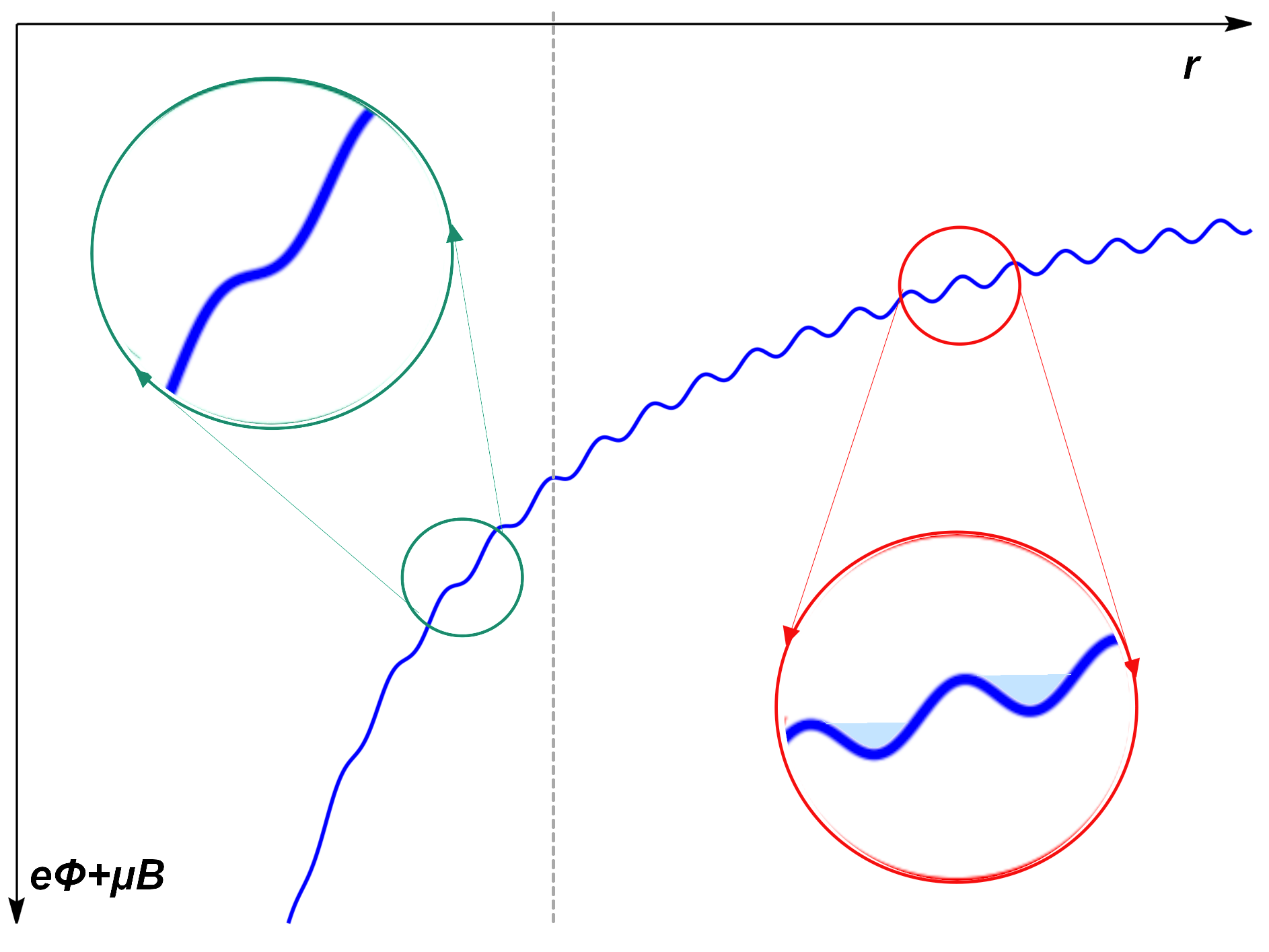}
        \caption{Radial dependence of the $e\Phi + \mu B$ effective potential, which governs electron motion. Magnetic field fluctuations are represented by a simple sinusoid. The vertical dashed line separates two radically different regions: on the left-hand side (closer to the Sun), the potential increases monotonically; on the right-hand side, local minima form particle traps.}
        \label{fig1}%
    \end{figure}

When an electron moves away from the Sun along a field line, it moves against the effective potential, and thus, its parallel velocity decreases as it loses kinetic energy. Similarly, if it approaches the Sun, it gains kinetic energy at the expense of the potential. An electron, which lacks sufficient parallel energy to escape the Sun entirely, will be reflected by the potential at some finite distance from the Sun. The parallel velocity of the electron is vanishingly small near the reflection point, and thus, even small variations of the effective potential can influence the parallel motion. On the other hand, the particle will keep most of its perpendicular velocity, since the latter only decreases together with the magnetic field magnitude, which changes relatively little if the reflection point is still in the inner heliosphere. Thus, the $\mu B$ contribution is significant in the energy balance.

For an electron to actually get trapped in an (effective) potential minimum, a kind of "viscosity" is needed, namely the particle should suffer some interaction, which can change its velocity quasi-randomly. Otherwise, the reflected particle, moving backward, can roll over every potential barrier it encountered during its outward travel. Coulomb collisions can cause significant alteration of the distribution on large scales \citep{horaites2019, boldyrev2019, boldyrev2020}, but near-Sun, wave particle interactions and resonant effects are more important \citep{horne2007, saito2007, thorne2010, kajdic2016, karbashewski2023, vo2024}. Particularly interesting for us are the results of \cite{tobita2018}, who have shown that when a particle interacting with a wave moves in a main potential slowly varying in space, the potential "drags the particle through" the resonant condition in velocity space, and the resonant wave-particle interaction causes significant alteration of its motion. According to their results, the alteration of the velocity in this resonant process is of the same order of magnitude as the velocity itself. In our case, every electron encounters the resonant condition before it is reflected by the solar electrostatic potential (when it is spatially close to the reflection point and its parallel speed approaches the wave propagation speed). We thus have a solid theoretical basis for concluding that the reflected particles form a population that is significantly scattered relative to the original, thereby satisfying the “viscosity” requirement. \cite{halekas2021} provides further support by reporting compelling observational evidence of processes that act to isotropize the electron velocity distribution function within a few tenths of an AU from the Sun. 

Outmoving electrons starting from and reflected inside the critical distance $r_c$ can move back towards the Sun (and the measurement site) unhindered. Even if the particle loses some of its parallel energy to resonant scattering, it will not encounter any potential walls during its inward motion. However, if an outmoving electron has sufficient parallel energy to reach the region beyond $r_c$, it will meet outward moving potential walls due to the moving magnetic traps. If the particle can lose some of its parallel energy to some of the above-mentioned processes, it will get trapped and move farther outward. Such particles will not return to the measurement site, and thus this process causes a sunward electron deficit above a certain energy level. The critical cutoff energy associated with this process is the potential difference ($\Delta \Phi$) between the measurement site and the critical distance. This difference is less than the true depth of the solar electric potential ($\Phi(r_0) = \Phi(r_c) + \Delta \Phi$); where  $\Phi(r_c)$ is the value of the potential at the critical distance. The location of the boundary is determined by the first local maximum of the effective potential because the first trap (first local minimum) forms behind that potential wall. We can quantify the relationship between the cutoff energy and the true depth of the potential using the critical distance. Supposing that the potential behaves as \(\Phi\sim\frac{1}{r^{\alpha}}\), we can relate the depth of the potential well (\(\Phi(r_0)\)) to the potential drop measured by the cutoff (\(\Delta\Phi\)) as:
\begin{equation}\frac{\Delta \Phi}{\Phi(r_0)} = 1- \Big(\frac{r_{0}}{r_c}\Big)^\alpha,
\label{DPhi}
\end{equation}
where \(r_{0}\) is the radial distance of the measurement site from the Sun.

Provided that the residual potential $\Phi(r_c)$ equals the energy deficit of solar wind acceleration – the additional energy needed to reach the observed solar wind speed but not accounted for by the electron cutoff – the critical distance can be determined. \cite{bercic2021} established that 59\% of the final kinetic energy can be explained by a potential proportional to the cutoff, which means that the average energy deficit is 41\%. From this we can immediately deduce that the critical distance is around $30 - 50 R_S$, since $r_c = \Big(\frac{\Phi(r_0)}{ \Phi(r_c)}\Big)^{1/\alpha} r_{0}$ for an electric potential decreasing as $1/r^\alpha$ with radial distance from the Sun. This, of course, is only a rough estimate of the average location of the boundary. Since conditions can vary from flux tube to flux tube, and electron deficit can sometimes be observed further from the Sun, $r_c$ is also highly variable and can be as much as 2-3 times larger for specific field lines.

Using the hypothesis that the measured cutoff equals the depth of the potential at the measurement site, \cite{bercic2021} fitted a simple power law on the data and found that the potential decreases as $\Phi_0/r^{0.66}$ between 20.3 and 85.3 $R_S$. If, however, the cutoff only represents part of the potential, an inhomogeneous function, $\Phi_0/r^{\alpha}-\Phi(r_c)$ should be fitted instead. Since this new function has an additional parameter, the problem is under-determined. However, if we have a priori knowledge of either $\alpha$ or $\Phi(r_c)$, we can easily compute the other parameter. Furthermore, we can establish that the presence of the residual potential requires higher $\alpha$ values compared to the $\Phi(r_c)=0$ ($r_c=\infty$) case. The higher the residual potential, the steeper the optimal power-law fit. The $\Phi_0/r^{0.66}$ function can, for example, be approximated within 5\% with the new inhomogeneous function using $\alpha=1$ if the residual potential ($\Phi(r_c)$) equals one third of the potential at $20.3 R_S$, and even $\alpha=1.33$ gives an acceptable result within 7.5\% of the original fit of \citep{bercic2021}. For a residual potential, which is 40\% of the total potential $\alpha=1.33$ already fits better, within about 5\% of the \citep{bercic2021} approximation. This finding favors the collisionless exospheric models, which predict a power law index $\alpha=1.33$ \citep{meyer-vernet1998, zouganelis2004}.

\section{Discussion}
Until now, we have used the general term "magnetic trap" to describe the phenomenon that interact with the electrons and sweep them away from the Sun. It is important to reach a better understanding of the nature of these magnetic traps, to be able to quantify their effects. 

To trap particles, the phenomena needs features that act like local potential wells. Electric fields can obviously provide such features, but magnetic fluctuations can also suffice, especially fluctuations, in which the field magnitude also changes. Since electrons travel along field lines, we are most interested in field line structure and fluctuations propagating along the field. However, measurements do not provide direct clues about these. The spacecraft trajectory  generally crosses flux tubes and field lines. Although there are rare instances in which the probe might follow a field line for a brief period, the fluctuations it measures typically reflect changes unrelated to our primary focus.
Even if the spacecraft moves in the magnetic field direction generally, the electron gyroradius is so small that the probe crosses back and forth between field lines, the perpendicular extent of a field line here practically defined by the gyromotion of the electrons. Having said that, \cite{zhao2021}  observed the wave activity in near-Sun field aligned flows. Although Alfvenic fluctuations dominated these intervals, they found strong evidence of outward-propagating fast magnetosonic waves. Purely Alfvenic waves  cannot entrap particles, although strong, nonlinear Alfenic waves feature field magnitude variations, and thus such waves can. Magnetosonic waves also have the potential to do so. However, the characteristic extent of the observed field magnitude variation in these waves is only a few percent of the total field, and thus the variation of the $\mu B$ part of the effective potential is also rather small. 

However, we can show that there is a type of fluctuation associated with much deeper potential wells. Magnetic switchbacks \citep{bale2019,kasper2019} are essentially Alfvenic structures, although sometimes there are field magnitude variations near their boundaries (for a review, see \cite{badman2026}). However, their importance as particle traps is due to another effect. True switchbacks, as the name suggests, feature a magnetic field direction turnover, and thus escaping electrons actually travel towards the Sun inside these structures. They represent a serpentine path for the electrons, on which the outward traveling electrons first turn back towards the Sun, and then again turn antisunward. On the sunward part of their path, the solar electric potential will accelerate them, contrary to the usual decelerating effect. Thus, the electrons will experience a potential well as they move along the field lines. 
Thus, it is plausible to consider switchbacks as significant actors in particle trapping. Below, we will discuss the effects of two kinds of magnetic disturbances, first the fluctuations of magnetic field magnitude, which act through the magnetic mirror force, and second the switchbacks.

If the traps act through magnetic mirror forces, the reflected
electron should be outside the loss-cone of the bottleneck of the magnetic trap, and the magnetic mirror force ($F_m = - \mu  \frac{\partial B}{\partial r}$)
should be equal to or greater than the parallel electrostatic force (\emph{eE}) to be able to compensate for the inward forcing effects of the electric field. 
The effects of the electrostatic force on the particle reflected near the critical distance can be expressed as:
\begin{equation}
eE =-e\frac{\partial\Phi}{\partial r} =\frac{e\alpha \Phi_0}{r_c^{\alpha+1}} = \frac{e\alpha \Phi(r)}{r_c} \approx \frac{e\Delta\Phi}{r_c-r_0} \approx  \frac{mv_{\parallel}^{2}}{2(r_c-r_0)},
\label{electric}
\end{equation}
where, for the last two estimations, we approximated the electric field by the finite difference quotient. On the other hand, from the definition of the mirror force we have:
\begin{equation}F_{m} = - \mu  \frac{\partial B}{\partial r} = - \frac{mv_{\bot}^{2}}{2B}\frac{\partial B}{\partial r} \approx \frac{mv_{\bot}^{2}B_1}{2\lambda B_0},\end{equation}
where $B_1$ and $\lambda$ are the characteristic magnitude and wavelength of magnetic fluctuations, respectively, and $B_0$ is the magnitude of the main field.
At the critical distance \({eE = F}_{m}\), which means that the critical distance depends on the perpendicular velocity of the particle as
\begin{equation}
    r_c \sim v_{\bot}^{-2/(\alpha+1)},
    \label{rc}
\end{equation}
because, as we mentioned earlier, since $\mu$ is only a constant of motion, the effective potential varies with the velocity of the particle. Comparing this result with Eq. \ref{DPhi}, we find that for magnetic mirror traps, the cutoff energy is highly dependent on the perpendicular velocity. However, the situation is more complex than suggested by Eq. \ref{rc}, as the ratio \(B_1/B_0\) can also vary with \(r_c\). Additionally, the perpendicular velocity \(v_\bot\) in the equation is evaluated at the critical distance, so it is also \(r_c\)  dependent. 

Despite these complexities, the overall understanding remains consistent: electrons with high perpendicular velocities will experience a lower cutoff, while those with negligible perpendicular velocity can be lost only if their parallel energy is sufficient to escape the total potential. Although the reflected electron population is significantly scattered compared to the original distribution, it may be worthwhile to investigate remnants of this strong initial dependency on perpendicular velocity within the reflected electron population.

Using the last approximation of Eq. \ref{electric}, the force balance for magnetic mirror traps can be re-stated as a loss-cone condition, a relationship that the ratio of parallel and perpendicular velocities of a trapped electron should satisfy.
The \({eE < F}_{m}\) condition is fulfilled if
\begin{equation}\frac{v_{\parallel}^{2}}{v_{\bot}^{2}} < \frac{(r_c-r_0)}{\lambda}\frac{B_1}{B_0}.
\label{loss-cone}
\end{equation}
The first quotient on the right-hand side represents the ratio of the distance between the reflection point and the measurement point to the wavelength of the fluctuations, while the second represents the weight of the fluctuations relative to the main field. 

In theory, we should also be able to determine $r_c$ from measurement data. To do this, we should only locate the first zero of the derivative of the effective potential, a.k.a the smallest $r$ for which the electric force equals the magnetic trapping force. This process is straightforward for a well-behaved function; however, since the power of solar wind magnetic field fluctuations is spread across a very wide frequency range, it is not easy to establish a "typical" amplitude for its derivative.

If we try to compute the derivative directly from the magnetic field measurements conducted during the time intervals analyzed by \cite{halekas2022}, we find that the result is primarily influenced by the highest frequency fluctuations. Recent results \citep{mcintyre2025} suggest that these high-frequency fluctuations are caused by ion cyclotron waves amplified by the helicity barrier effect \citep{meyrand2021}. These fluctuations, however, correspond to very shallow dents in the effective potential easily passed by even the thermal electrons. We can determine a typical value for the derivative of larger fluctuations, which is approximately $10^{-4} - 10^{-5}$ nT/km provided that the velocity of the fluctuations in the spacecraft frame is within a factor of two of the solar wind speed. Depending on the magnetic moment, this translates to \(\mu \frac{\partial B}{\partial r} = 10 - 1000 \, \text{eV/Gm}\). This range is plausible since \(eE\) is of a similar order of magnitude and therefore they may balance each other out. However, we must conclude that the measurements are not yet accurate enough to allow a direct computation of the critical distance.

Next, we analyze the magnetic traps formed by switchback-related bends of the field lines. The depth of these potential wells can be approximated from the length (radial extent) of the switchbacks, their deflection angles, and the local value of the solar electrostatic field. The last proportionality seems to suggest that there is no radial evolution of the force balance, which could define a barrier separating distinct regions of returning and escaping particles as depicted in Fig. 1. As the electric field decreases, so does the depth of the local potential wells, which seemingly prevents the formation of the barrier. This simple argument, however, does not account for the radial evolution of the switchback deflection angle. \cite{jagarlamudi2023} has shown that the relative frequency of high deflection angle "true" switchbacks (in which the electrons really turn back towards the Sun) increases almost an order of magnitude between 20 and 40 $R_S$. This suggests that low-deflection disturbances evolve into high-deflection switchbacks in this radial region. While the evolution of individual switchbacks is a continuous process, the formation of an effective trapping barrier is statistically sharp due to the exponential radial evolution of the their occurrence rate. Since switchbacks only appear as local potential minima if their deflection angle exceeds $90^{\circ}$, this evolution produces the same qualitative behavior illustrated in Fig. 1. The critical distance $r_c$ can be identified by the position where the relative frequency of high deflection switchbacks reaches its maximum, which is between 30 and 50 $R_S$. This result agrees well with our earlier approximation of $r_c$ based on the observed energy deficit of solar wind acceleration.

The strength of the electric field can be estimated from the cutoff and the critical distance using the finite differentia quotient approximation as shown in Eq. \ref{electric}. We find that the electric field is around 5 nV/m.

\cite{horbury2020} found that switchbacks are elongated structures, their widths are around $10^4$ km, and their length is at least several times larger, suggesting that their length scale is on the order of $10^5$ km. \cite{pecora2022} determined the spatial extent and frequency of switchbacks. Their results also suggest $10^5$ km for the average length of these structures, while their characteristic separation is $10^6$ km. \cite{laker2021} has found that the typical width of the structures is $5\times 10^4$ km, with an aspect ratio of the order of 10. Combining these results with the estimate for the electric field, we find that the potential drop inside a $180^\circ$ switchback is on the order of 1V.


\cite{jagarlamudi2023} have found that large amplitude switchbacks are much less frequent in the slow solar wind than in the fast wind. We can use this result to explain the findings of \cite{halekas2022} that the observed energy cutoff (interpreted as the depth of the total potential) can only account for the acceleration of the slowest wind. Since large switchbacks are rare in the slowest wind, there are no moving traps to carry away those electrons, which do not have enough energy to escape the solar electric potential entirely. Thus, the cutoff is comparable to the total depth of the potential there. However, in the fast wind, the switchbacks work as a conveyor belt for "tired" electrons, and thus the cutoff represents only a local potential difference in accordance with Eq. \ref{DPhi}.

Another important aspect of the above results concerns the core population of the electron component of the solar wind. It is often assumed that this component consists of electrons trapped between the outer electrostatic and the inner magnetic mirror reflection points \citep[see e.g.:][]{boldyrev2020,lie-svendsen2000,marsch2006,meyer-vernet1998}. However, these reflection points are considered global in the sense that the magnetic mirror arises from the global divergence of the main magnetic field lines (not the fluctuations), and the electrostatic reflection point is determined by the global properties of the electrostatic field without any influence from magnetic fluctuations. \cite{boldyrev2020} explains the observation that the core electron population moves at a speed close to that of the solar wind with a slow collisional energization of this trapped population. 
Here, we have presented another process that may contribute to the formation of the core population. As the decelerated electrons are caught in the traps formed by magnetic fluctuations, they contribute to a low-energy electron population. This population moves together with the traps. Since the magnetic traps travel at a speed close to that of the solar wind, this low-energy population moves in tandem with the ion component, as expected from the electron core. Since magnetic trap characteristics are likely to vary from one flux tube to another, this mechanism may help explain the significant changes observed in core electron properties at flux tube boundaries \citep{borovsky2021}.

It is also important to consider in this context that the average speed of electrons reflected inside $r_c$ will also be affected by the motion of magnetic disturbances. Fluctuations in this region cannot establish stable magnetic traps characterized by minima in the effective potential, but they can create moving enhancements of the potential gradient (cf. Fig. 1, the situation inside $r_c$). Electrons are likely to be reflected by these moving "walls". This phenomenon occurs at both the outer and inner reflection points. At the outer reflection point, an outward-moving electron is reflected by an outward-moving wall, resulting in an inward-moving electron that is slower by twice the speed of the moving fluctuation. Conversely, at the inner reflection point, the situation is reversed: the reflected electron gains speed that is twice that of the magnetic fluctuations there. Overall, this results in a shift in the average speed of the outward-moving electron population compared to the inward-moving population. Consequently, the average speed of the electrons aligns with that of the moving enhancements of the effective potential gradient corresponding to magnetic fluctuations. If the speed at the outer reflection point exceeds that at the inner reflection point (due to solar wind acceleration, for example), this process also contributes to electron cooling.

\section{Conclusions}
We have shown that the cutoff energy observed by the Parker Solar Probe in the sunward moving portion of the electron distributions close to the Sun is not necessarily directly related to the depth of the electrostatic potential well at the measurement site, instead it characterizes the potential drop inside an $r<r_c$ region around the Sun, where the contribution of magnetic field fluctuations to the effective potential governing the parallel motion of the electrons cannot yet establish stable magnetic traps characterized by minima in the effective potential. Particles that escape this region will not be seen again by the space probe, because they will be caught and carried away by magnetic traps. 

As a result, the sunward-moving electrons detected by PSP are those already reflected inside $r_c$, and thus their energy cannot exceed the potential drop inside this region. Hence, the observed energy cutoff is proportional to this potential drop ($\Delta\Phi$). The ratio of the cutoff to the depth of the potential well ($e\Phi(r_0)$) is: $\frac{\Delta \Phi}{\Phi(r_0)} = 1- \Big(\frac{r_{0}}{r_c}\Big)^\alpha$, where $r_0$ is the radial distance of the measurement site from the Sun, $r_c$ is the critical distance where the trapping begins, and $\alpha$ is the exponent characterizing the radial behavior of the solar electrostatic potential. 

This means that the actual depth of the potential well can be significantly larger than that computed from the observed cutoff. Consequently, the contribution of the electric potential to solar wind acceleration can be more important than prior interpretations of the cutoff would suggest. Conversely, the critical distance can be estimated from the cutoff value, assuming that the true depth of the potential equals the energy needed to accelerate the ions to the observed solar wind speed. From this, we can conclude that $r_c$ is about $30 - 50 R_S$. 

We have found that in the presence of a finite $r_c$, the radial variation of the solar electric potential requires a steeper power-law fit, and thus the value of the exponent $\alpha$ may be close to the $\alpha=1.33$ value required by collisionless exospheric models. 

We examined two types of magnetic disturbances that can form moving local traps: magnetic field intensity variations and magnetic switchbacks. 

For intensity variations, the critical distance, and therefore the cutoff, is not universal; specifically, they depend on the perpendicular velocity of the particle. Although the reflected electron population is significantly scattered compared to the original distribution, it may be worthwhile to investigate remnants of this strong initial dependency on perpendicular velocity within the reflected electron population. 

In the case of switchbacks, $r_c$ and thus the cutoff are universal. The critical distance can be identified as the radial position where the relative frequency of high-deflection switchbacks reaches its maximum, which occurs between 30 and 50 $R_S$. This finding aligns well with our earlier approximation of $r_c$ based on the observed energy deficit of solar wind acceleration. 

Since switchbacks are much less frequent in the slow solar wind compared to the fast solar wind, there are no moving traps in the slow wind to carry away electrons that do not have enough energy to escape the solar electric potential completely. Consequently, in the slow wind, the cutoff is comparable to the total depth of the potential, while, in contrast, in the fast wind, the cutoff represents only a portion of the accelerating potential. This agrees well with the findings of Halekas et al. (2022), which indicated that the observed energy cutoff (interpreted as the depth of the total potential) can only account for the acceleration of the slowest solar wind.

\bibliographystyle{bibtex/aa} 
\bibliography{Complete_SW_Lib_BB} 
\end{document}